\documentclass[prb,twocolumn,amsmath,amssymb,bibnotes]{revtex4}
\usepackage{graphicx,epsfig}
\usepackage{graphicx}% Include figure files
\usepackage{dcolumn}% Align table columns on decimal point
\usepackage{bm}% bold math
\usepackage{html,makeidx}

\begin{document}

%\preprint{APS/123-QED}

\title{Scattering Mechanism in Modulation-Doped Shallow Two-Dimensional Electron Gases}

\author{D. Laroche$^{1,2}$, S. Das Sarma$^{3}$, G. Gervais$^{2}$, M. P. Lilly$^{1}$, and J. L. Reno$^{1}$}
\address{$^{1}$Center for  Integrated Nanotechnologies, Sandia National Laboratories, Albuquerque, NM 87185 USA}
\address{$^{2}$Department of Physics, McGill University, Montreal, H3A 2T8 CANADA}
\address{$^{3}$Condensed Matter Theory Center, Department of Physics,
University of Maryland, College Park, MD 20742-4111 USA}

\date{\today}

\begin{abstract}
We report on a systematic investigation of the dominant scattering mechanism in shallow two-dimensional electron gases (2DEGs) formed in modulation-doped GaAs/Al$_{x}$Ga$_{1-x}$As heterostructures. The power-law exponent of the electron mobility versus density, $\mu\propto n^{\alpha}$, is extracted as a function of the 2DEG's depth. When shallower than 130 nm from the surface, the power-law exponent of the 2DEG, as well as the mobility, drops from $\alpha \simeq1.65$ (130 nm deep) to $\alpha \simeq1.3$ (60 nm deep). Our results for shallow 2DEGs are consistent with theoretical expectations for scattering by remote dopants, in contrast to the mobility-limiting background charged impurities of deeper heterostructures.
\end{abstract}

%\pacs{Valid PACS appear here}  % PACS, the Physics and Astronomy
                               % Classification Scheme.
%\keywords{Suggested keywords} %Use showkeys class option if keyword
                               %display desired
\maketitle

Ever since the development of modulation doping \cite{Stormer-doping1} in the early 1980's, the quality
of  GaAs/AlGaAs semiconductor heterostructures  grown by molecular beam epitaxy
has been steadily improving. With this technique, the mobility of electrons forming a  two-dimensional electron gas (2DEG)
at semiconductor interfaces can now exceed $\sim 3 \times 10^{7}$ $cm^{2}/(V\cdot s)$.  In such high mobility structures\cite{Pfeiffer}, transport is usually dominated by scattering from the unintentional background charged impurities since the doping layers, being very far from the 2DEG, do not contribute significantly to the resistivity.  The large separation between the 2DEG and the doping layers ($\sim 100$ $nm$) leads to a large separation between the 2DEG and the surface and makes such heterostructures less than ideal for the patterning of nanostructures.  Indeed, this large separation causes the nanostructure's confinement potential created by depletion top gates and/or by wet etching to be less abrupt, complicating the fabrication of the smallest nanostructures.  This limitation can be overcome  by working with undoped heterostructures where the absence of doping layers allows the formation of shallow 2DEGs with relatively high mobilities \cite{Kane1, Kane2, Lilly, Herfort, Frost, Holland, Saku1, Saku2}.  The drawback of this approach, however,  is that undoped structures usually require more complicated processing, {\it e.g.} they often require  an additional accumulation gate and the use of overlap or self-aligned contacts \cite{Kane1}. These additional processing steps make the fabrication of laterally \cite{Peta} and vertically-coupled nanostructures \cite{Ed} significantly harder to realize. For this reason, the refinement of growth techniques and even more so a clear understanding of the scattering mechanism in shallow 2DEG appear crucial to the development of coupled nanostructures with low-disorder.

In this work, we present an investigation of the dominant scattering mechanism in shallow GaAs/AlGaAs modulation-doped 2DEGs with quantum wells grown systematically closer to the surface.  Such structures allow easier fabrication processing while keeping the enhanced disorder from the doping layers at a reasonably low level.  To extract the dominant scattering mechanism, we measure the density dependence of the 2DEG mobility using gated Hall bars.  This provides important information on the mobility-limiting mechanisms in shallow modulation-doped 2DEGs,  and also provides useful insights for future modeling of shallow heterostructures.

All samples described in this Letter are GaAs/Al$_{x}$Ga$_{1-x}$As heterostructures grown by molecular beam epitaxy with the generic structure sketched in Figure 1, panel (a).  Epilayers were grown on an undoped (100) GaAs substrate.  At first, a GaAs buffer, an Al$_{0.55}$Ga$_{0.45}$As superlattice and an Al$_{0.24}$Ga$_{0.76}$As spacer were grown.  One or two doping layers, depending on the depth of the structure, were then added.  In the two doping layers case, an Al$_{0.24}$Ga$_{0.76}$As spacer was added between both layers.  Additionally, an Al$_{0.24}$Ga$_{0.76}$As setback layer, a 30 nm GaAs quantum well and another setback layer were grown. Finally, one or two additional doping layers and their respective spacers (as well as a GaAs cap) were added on the top of the structure.    The density of dopants, $n_{\delta}$, was  kept constant for every doping layer in a given heterostructure.  The doping layer positions, labeled $d_{1}$, $d_{2}$, $d_{3}$ and $d_{4}$ going from the top to the bottom of the heterostructure, are determined from the closest quantum well interface.  The depth of the sample is defined as the distance between the top surface and the middle of the quantum well.  The location and density of the doping layers, as well as the sample depth, are given in Table 1 for every heterostructure used in this work.  The position and density of the dopants were selected to produce an ungated 2DEG density between $\sim 1 \times 10^{11} cm^{-2}$ and $\sim 3 \times 10^{11} cm^{-2}$.  Then, for each depth, heterostructures with various dopants position and density were grown with the ones yielding the best ungated mobility being selected.  Note that as the depth of the quantum well is reduced, it is necessary to move the delta-doping layers closer to the quantum well in order to keep the 2DEG density in the desirable range.  This process was performed both for symmetric and asymmetric positioning of the doping layers around the quantum well.

\begin{figure}
\begin{center}
\includegraphics[width = 7.5cm]{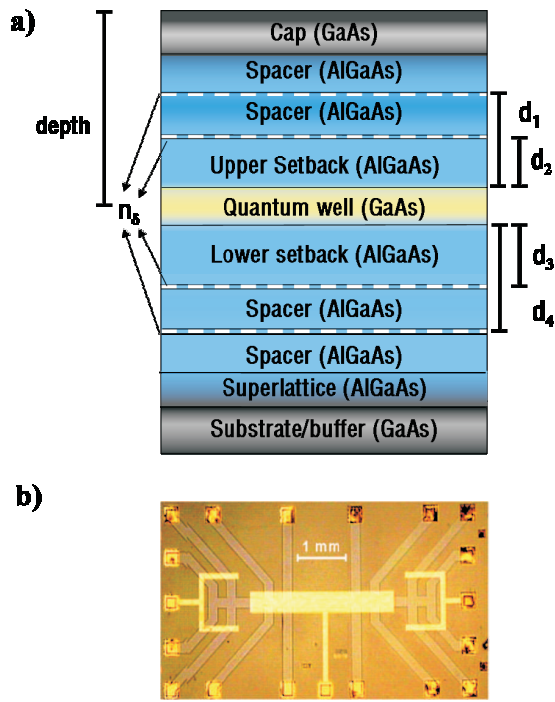} \\
  \caption{\label{fig0} \textbf{(a)} Sketch of the generic modulation-doped GaAs/AlGaAs heterostructures, with the doping layers shown as dotted lines.  The depth of the 2DEG is measured from the top of the heterostructure to the middle of the quantum well. The density of the dopants is $n_{\delta}$ and  $d_{i}$ denotes the distance between the $i^{th}$ doping layer and the closest interface of the quantum well. \textbf{(b)} Photograph of a typical device showing the gated Hall bar pattern used in this experiment. }\vspace*{-1mm}
\end{center}
\end{figure}

\normalsize{}
In order to measure the electron mobility and vary the electronic density, a standard TiAu-gated Hall bar was patterned on top of the heterostructures.  A photograph of a typical structure is shown in Figure 1, panel (b).  The resistivity was measured using a low-frequency standard measurement technique with a constant excitation current of 10 nA at 9 Hz.  Measurements were performed at a temperature of 4 K in a magnetic field ranging from 0 to 0.172 T. The electron density was determined by measuring the slope of the low magnetic field Hall resistance ($R_{xy}$).  Independent measurements of the density on selected heterostructures using the minima of the $\rho_{xx}$ oscillations at 0.3 K proved to be identical within error to those extracted from the slope of the low magnetic field Hall resistance.

\begin{figure}
\includegraphics[width = 9cm]{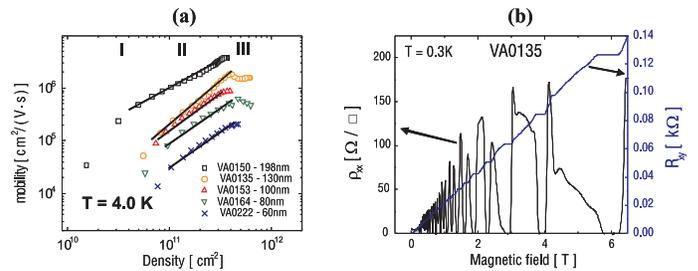}
\vspace*{-1mm}
\caption{\label{fig0} \textbf{(a)} Log-log plot of the mobility versus the electronic density for selected heterostructures. The intermediate density regime is denoted by region II. The power-law fits in this region are shown as dotted lines. \textbf{(b)} Longitudinal resistivity (left axis) and Hall resistance (right axis) for the VA0135 sample.}
\vspace*{-1mm}
\end{figure}
\normalsize{}

The mobility measurements versus the electronic density are presented in Figure 2, panel (a) as a log-log plot.  Only selected heterostructures are shown for clarity, but similar results were obtained for all structures.  Three distinct mobility regimes are observed.  The regime of interest in this work is the intermediate density regime, shown as region II, where inter-subband scattering, density inhomogeneity and localization effects are minimal.  In this regime, the mobility ($\mu$) approximately scales with density ($n$) as $ \mu \propto n^{\alpha} $.  The power-law exponent $\alpha$ can be used to obtain information on  the dominant scattering mechanism affecting the 2DEG.  Several experimental and theoretical studies have used similar techniques to analyze disorder in 2DEGs \cite{Tsui, Jiang, Kane2, Saku1, Saku2, Ritchie1, Ritchie2, Percolation2, Das Sarma}, as recently discussed in Ref. 18.  For the lowest electronic density, region I, the mobility decreases more rapidly with decreasing density than in region II.  This low density regime is dominated by a fluctuation-induced density inhomogeneity eventually leading to a percolation localization transition to an insulating phase\cite{Percolation1, Percolation2} at a critical density $n_{c}\lesssim 5 \times 10^{10} cm^{-2}$ (for our samples).  Region III, the highest electronic density regime, is characterized by a mobility drop with increasing density.  This is a consequence of inter-subband scattering due to the population of the second energy subband of the quantum well.  This interpretation was confirmed by our investigation of the Shubnikov-De Haas oscillations at $T\simeq 0.3$ K in a high magnetic field (not shown in this Letter), which proved consistent with the occupation of a second subband.  Measurements of $\rho_{xx}$ and $R_{xy}$ were also performed in every heterostructure in the intermediate density regime at $T\simeq 0.3$ K and in high magnetic fields in order to determine whether or not parallel conduction played a role in our devices.  The $\rho_{xx}$ and $R_{xy}$ curves were similar in every heterostructure and a typical set (Va0135) is shown in panel (b) of Figure 2.  For all samples, the $R_{xy}$ trace is linear with magnetic field until the onset of typical integer Hall effect plateaus was reached.  Those plateaus are consistent with the minima in $\rho_{xx}$ trace, which are reaching a zero value of resistance within $0.2\%$ of the Lock-In amplifier full scale.  These results strongly suggest that parallel conduction in our devices did not play a significant role.  In addition, we performed a self-consistent Schr\"odinger-Poisson simulation on every heterostructures used in this Letter.  Assuming that the Fermi energy is pinned at mid-gap at the surface of the structures (cap) and that the electric field is zero at the bottom of the structures (superlattice), we found that electrons solely collect in the quantum well, consistent with an absence of parallel conduction in our devices.
\begin{table}
\small{}
\scalebox{0.8}{
\renewcommand{\arraystretch}{1.1}
\begin{tabular}{ c  c  c  c  c  c  c  c }

\textbf{Sample} & \textbf{Depth}  & $\large{\bm{n_{\delta}}}$ & $\large{\bm{d_{1}}}$ & $\large{\bm{d_{2}}}$ & $\large{\bm{d_{3}}}$ & $\large{\bm{d_{4}}}$ & $\large{\bm{\alpha}}$ \\
VA0150 & 198 nm & $1 \times 10^{12} cm^{-2}$ & ----- &  75 nm & 95 nm & ----- & $1.01 \pm 0.04 $ \\ \hline
VA0123 & 160 nm & $8 \times 10^{11} cm^{-2}$ & ----- & 75 nm & 75 nm & ----- & $1.25 \pm 0.03 $ \\ \hline
VA0135 & 130 nm & $1 \times 10^{12} cm^{-2}$ & 85 nm & 65 nm & 65 nm & ----- & $1.65 \pm 0.1$ \\ \hline
VA0142 & 100 nm & $2 \times 10^{12} cm^{-2}$ & 65 nm & 55 nm & 55 nm & 65 nm & $1.31 \pm 0.02 $ \\ \hline
VA0153 & 100 nm & $2 \times 10^{12} cm^{-2}$ & 65 nm & 55 nm & 75 nm & 85 nm & $1.31 \pm 0.03 $ \\ \hline
VA0161 & 80 nm & $2 \times 10^{12} cm^{-2}$ & 52 nm & 45 nm & 45 nm & 55 nm & $1.35 \pm 0.04 $ \\ \hline
VA0164 & 80 nm & $2 \times 10^{12} cm^{-2}$ & 50 nm & 42 nm & 62 nm & 72 nm & $1.28 \pm 0.03 $ \\ \hline
VA0222 & 60 nm & $3 \times 10^{12} cm^{-2}$ & 34 nm & 28 nm & 48 nm & 58 nm & $1.26 \pm 0.06 $ \\ \hline
\end{tabular}}
\caption{Parameters used in the growth of the GaAs/AlGaAs heterostructures. The depth of the quantum well, dopants density and dopants position, as defined in Fig.1, as well as the power-law exponent $\alpha$ are given for each sample.}\vspace*{-6mm}
\end{table}

Focusing on the intermediate density regime where the mobility is a power-law function of the electronic density, we extract the exponent $\alpha$ for each symmetrically- (open squares) and asymmetrically-doped (open circles) heterostructure, where the error bars on the data are accounting for the uncertainty on the intermediate density regime range.  The exponents are shown in Figure 3 versus the 2DEG depth and in Table 1.   Additional measurements performed on different samples originating from the same wafer yielded results within the error bars quoted in Table 1.  Our standard heterostructure with a 198 nm deep 2DEG (VA0150) is shown in the far right with a fitted exponent very close to unity, $\alpha=1.01\pm  0.04$.  Previous theoretical and experimental studies have shown that $\alpha \lesssim 1$ is expected as a trademark of the background unintentional charged impurities acting as the most important scattering source\cite{Tsui, Jiang, Das Sarma}.  For the 130 nm deep structure (VA0135), the power-law exponent rises to a value of $1.65\pm 0.1 $. This is indicative of scattering dominated by the unscreened remote dopants located in the doping layers \cite{Das Sarma}.  For heterostructures  grown with 2DEGs  shallower  than 130 nm from the surface, $\alpha$, as well as $\mu$ itself, decreases and reaches approximately $\simeq 1.35$ for symmetrically-doped  and $\simeq 1.3$ for asymmetrically-doped structures.  Ionization fraction of the dopants in the doping layers should only provide an overall scaling factor to $\rho$ or $\mu$, thereby not modifying the power-law exponent\cite{Das Sarma}.  Thus, considering the ionized impurity profile fixed in a single sample, the exact value of the ionization fraction in each sample should not play a significant role in this experiment.

\begin{figure}
\includegraphics[width = 7.5cm]{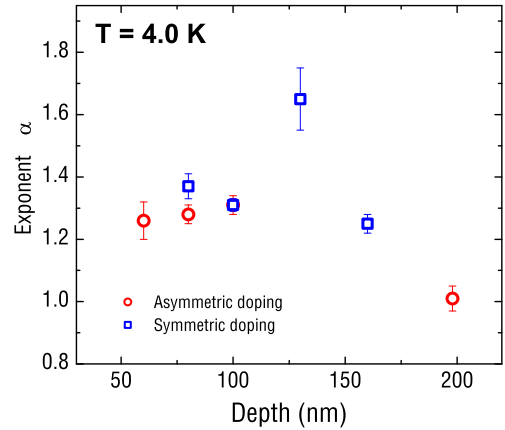}
\vspace*{-6mm}
\caption{\label{fig0} Power-law exponent $\alpha$ versus quantum-well depth for symmetrically-doped (open squares) and asymmetrically-doped (open circles) heterostructures.} \vspace*{-5mm}
\end{figure}
\normalsize{}

In general, one expects the mobility to be limited by a number of scattering mechanisms: background charged impurities, ionized dopants in the delta-doping layers, acoustic phonons, interface roughness and alloy disorder.  Scattering by optical phonons, which is the dominant resistive scattering mechanism from  room temperature down to $\sim 100$ K, is unimportant at 4 K in GaAs structures where the optical phonon energy is $\sim 35$ meV.  Since scattering from phonons, interface roughness and alloy disorder are expected to be weak \cite{Das Sarma} and unlikely to vary much with the depth of the 2DEG, we consider background impurities and remote dopants scattering to be the main resistive mechanisms in our samples.  This is consistent with our experimental findings with the highest mobility being achieved in our deepest sample (VA0150) where the remote dopants are presumably contributing little to the resistivity, the delta doping layers being farther from the 2DEG.  We can therefore assume that the dominant scattering mechanism in the highest mobility sample, VA0150, is the background impurity scattering, which is consistent with our exponent $\alpha \simeq 1$ as expected theoretically \cite{Das Sarma}.  As the 2DEG is moved closer to the surface, the modulation delta dopants are moved closer to the 2DEG to keep the ungated density roughly constant, leading to a stronger component of the scattering arising from the remote ionized impurities.  Samples shallower than 130 nm have lower $\mu$ (higher $\alpha$) than VA0150, the deepest sample.  It is therefore likely that these samples are dominated by remote dopants scattering.  This is also consistent with theory since $ \alpha \simeq 1.3$ in all these samples and manifests the tendency of decreasing $\alpha$ with decreasing spacer thickness as predicted theoretically \cite{Das Sarma}.  We therefore conclude that, except for VA0150 which is dominated by background impurity scattering, all shallower samples are strongly affected by remote dopant scattering.

In conclusion, we have fabricated doped GaAs/AlGaAs heterostructures with their quantum well located at variable depth and measured the mobility versus the electronic density curves.  The doping layers were moved closer to the quantum well as the structures became shallower in order to keep the ungated electron density roughly constant in the 2DEG.  In the intermediate density regime, where the mobility is approximately a power-law function of the electronic density, the extracted exponent shows a significant decrease when the 2DEG is grown closer than 130 nm from the surface.  This observation is qualitatively consistent with recent calculations by Hwang and Das Sarma for which the mobility-limiting mechanism was attributed to the remote dopants in the doping layers, and the decrease in the power-law exponent was attributed to increased proximity between the dopants and the 2DEG.  Since the structures used in this experiment were using large area surface gates, similar gate patterning techniques could easily be used to form nanostructures such as quantum point contacts or quantum dots.

This work was performed, in part, at the Center for Integrated Nanotechnologies, a U.S. Department of Energy, Office of Basic Energy Sciences user facility.  Sandia is a multiprogram laboratory operated by Sandia Corporation, a Lockheed Martin Company, for the United States Department of Energy under Contract No. DE-AC04-94AL85000.  It was also supported by the Natural Sciences and Engineering Research Council of Canada (NSERC).

\end{document}